
\input phyzzx.tex
\overfullrule=0pt
\tolerance=5000
\overfullrule=0pt
\twelvepoint

\twelvepoint
\pubnum{IASSNS-HEP-93/7}
\date{March, 1993}
\titlepage
\title{ROLE OF SHORT DISTANCE BEHAVIOR IN OFF-SHELL OPEN-STRING
FIELD THEORY}
\vglue-.25in
\author{Keke Li
\foot{W.M. Keck Foundation Fellow} and Edward Witten\foot{Research
supported in part by NSF Grant PHY91-06210.}}
\medskip
\address{School of Natural Sciences
\break Institute for Advanced Study
\break Olden Lane
\break Princeton, NJ 08540}
\bigskip
\abstract{A recent proposal for a background independent open string field
theory is studied in detail for a class of backgrounds that correspond to
general quadratic boundary interactions on the world-sheet. A short-distance
cut-off is introduced to formulate  the theory with a finite number of local
and potentially unrenormalizable boundary couplings. It is shown that
renormalization of the boundary couplings makes both the world-sheet partition
function and  the string field action finite and cut-off independent, although
the resulting string field action has an unpalatable dependence on the leading
unrenormalizable coupling.}
\endpage

\def\half{{1\over 2}}

\def\al{\alpha}

\def\Si{\Sigma}

\def\f{\phi}

\def\th{\theta}

\def\ep{\epsilon}
\def\de{\delta}
\def\De{\Delta}

\def\pa{\partial}

\def\la{\langle}
\def\ra{\rangle}

\def\e{{\rm e}}

\def\factor{{1\over 8\pi}}
\def\brst{Q_{\rm BRST}}
\def\tr{{\rm Tr}}
\def\mn{{\mu\nu}}
\def\nm{{\nu\mu}}
\def\sumkp{\sum_{k=1}^\infty}
\def\sumk{\sum_{k\in Z}}
\def\sumn{\sum_{r=0}^{s}}

\def\emkt{\e^{-ik\th}}
\def\ekep{\e^{-k\ep}}
\def\ept{$\ep\to0$}
\def\da{$\De a$}

\def\IR{{\hbox{{\rm I}\kern-.2em\hbox{\rm R}}}}
\def\IB{{\hbox{{\rm I}\kern-.2em\hbox{\rm B}}}}
\def\IN{{\hbox{{\rm I}\kern-.2em\hbox{\rm N}}}}
\def\IC{{\ \hbox{{\rm I}\kern-.6em\hbox{\bf C}}}}

\def\IZ{{\hbox{{\rm Z}\kern-.4em\hbox{\rm Z}}}}
\def\to{\rightarrow}
\def\d{{\rm d}}
\def\underarrow#1{\vbox{\ialign{##\crcr$\hfil\displaystyle
{#1}\hfil$\crcr\noalign{\kern1pt
\nointerlineskip}$\longrightarrow$\crcr}}}
\def\ltorder{\mathrel{\raise.3ex\hbox{$<$}\mkern-14mu
             \lower0.6ex\hbox{$\sim$}}}
\def\lesssim{\mathrel{\raise.3ex\hbox{$<$}\mkern-14mu
             \lower0.6ex\hbox{$\sim$}}}

\chapter{Introduction}

\REF\w{E. Witten, {\it On Background Independent Open-String
Field Theory}, Phys. Rev. {\bf D46} (1992) 5467.}
\REF\ww{E. Witten, {\it Some Computations in Background Independent Off-Shell
String Theory}, IASSNS/HEP-92/63.}

Recently, a new open string field action has been proposed using the
Batalin-Vilkovisky formalism[\w]. Formally, this action, defined in the space
of all world-sheet open string theories, is background independent and gives
the expected classical field equations and on-shell gauge symmetry.

It has been emphasized in [\w] that the construction of the string field
action has been formal because ultraviolet divergences associated with
unrenormalizable local world-sheet interaction have been ignored.
This question is crucial because the generic world-sheet theory is
unrenormalizable; the massive states of the string are represented in the
world-sheet Lagrangian by unrenormalizable interactions, as are high frequency
modes of massless states.  The difficulty in making sense of the generic
two dimensional Lagrangian has indeed long been one of the main obstacles
to progress in string theory; it severely limits applicability of
the world-sheet approach to string theory.

Since it is hard to find a general way
to remove the cut-off, one can simply define
the string field action in a space of cut-off interactions.  If the cut-off is
strong enough, there appears to be no problem with any of the considerations
of [\w]. However, the expected classical solutions of the theory probably
cannot be found in a space of world-sheet theories with strongly cut-off
boundary interactions (since the standard perturbations are by dimension one
local operators on the boundary which are not ``soft''); and a space of such
interactions probably cannot be defined in a background independent way
(the obvious way to get a space of strongly cut-off theories is to start
with a particular background, which cannot be ``soft'' since no theory is,
and then add to it a family of soft perturbations).  So it does not
appear adequate to define the string field action only in a space of strongly
cut-off interactions. However, one can begin by defining it in such a space
and then try to remove the cut-off.  In fact, that is what we will do in this
paper.

The difficulties with unrenormalizable theories arise because the short
distance behavior is out of control and depends on the specific Lagrangian;
and therefore the world-sheet anomalies, which are so important in string
theory, are also out of control.  To probe these issues, we will
consider a family of free field theories with quadratic but higher
derivative boundary couplings; being free, these theories are tractable,
but the short distance behavior depends on the specific couplings. For
motivation, we first recall the example considered in [\ww]; the bulk action
was the standard closed string action
$$L_{\rm I}=\int_\Sigma\,\d^2\sigma\sqrt{h}\left({1\over 8\pi}h^{\alpha\beta}
\partial_\alpha X_\mu\partial_\beta X_\nu\,\eta^\mn
+{1\over 2\pi}b^{\alpha\beta}D_\alpha c_\beta\right),         \eqn\L$$
and the boundary action was
$$ L'_{\rm B}=\oint_{\partial\Sigma}\d\theta \left({a\over 2\pi}
+\sum_{\mu=1}^{26}{u^\mu\over 8\pi}X_\mu{}^2 \right)\, .\eqn\Lp$$
Here the world-sheet is   a disc $\Si$ with unit radius, and $\{a, u^\mu\}$ is
a set of parameters. The boundary interaction represents a quadratic tachyon
background and is super-renomalizable; thus it only requires proper
normal-ordering (corresponding to absorbing an infinite constant into $a$) to
make both the world-sheet partition function and string field action
well-defined.

The discussion in this paper will follow that of [\ww] closely, but now with
the most general quadratic boundary action:
$$ L_{\rm B}= a + \factor\oint\d\th\d\th'X_\mu(\th)\,u^\mn(\th-\th')
X_\nu(\th')\, .\eqn\Lb$$
At this stage we need to decide what kind of function $u$ is to be.
The only evident notion of locality is that $u$ should be a finite
sum of derivatives of delta functions:
$$ u^\mn(\theta-\theta')=\sum^s_{r=0} t^\mn_r{d^r\over d
\theta^r}\delta(\theta-\theta')\, ,  \eqn\mmc$$
with $t_r^\mn$ being constants. With such a choice, most axioms of local
quantum field theory are preserved
(but world-sheet unitarity is lost because of the higher derivatives); the
short distance behavior of the propagator depends on $s$, leading to some
unpleasant properties that we will see later.  Alternatively, one can try
to take $s\to \infty$; this would even appear to be dictated by the fact
that, once massive fields are excited in string theory, one should expect
fields of arbitrarily high mass to be excited.  But with $s=\infty$, the sum
in \mmc\ is no longer local.
For instance, the ``identity''
$$X(\theta)X(\theta+\epsilon)=\sum_{n=0}^\infty {\epsilon^n\over n!}
X(\theta){d^n\over d\theta^n}X(\theta) \eqn\cmmmc$$
shows that any bilocal expression can be expanded formally as an infinite
sum of local operators.
What kind of function we get upon taking $s\to\infty$ in
\mmc\ depends on what large $r$ behavior we assume for
$t^\mn_r$.  For instance, a suitable condition
on the $t^\mn_r$ would give a class of strongly cut-off boundary interactions,
as discussed above.  One of the basic puzzles about our problem is that
apart from the local interactions (finite $s$) which have their own
difficulties that we have sketched, we do not know a natural class
of boundary interactions to focus on.

The quadratic nature of the boundary interactions makes the world-sheet theory
exactly soluble, even though the  dependence of the short-distance behavior on
the Lagrangian would usually be characteristic of unrenormalizable theories.
This will be discussed in Section 2, where the exact partition function is
determined from the exact Green's function. In practice, our way of grappling
with the issues introduced above will be to  introduce a regulator
corresponding to a boundary cut-off $\ep$ that replaces $\de(\th-\th')$ by
$\de_\ep(\th-\th')$
with $\lim_{\ep\to 0}\de_\ep(\th-\th') = \de(\th-\th')$.
The regularized theory thus has a smooth coupling function $u^\mn(\th-\th')$
and a non-local interaction \Lb. After computing the partition function and the
string field action, we will then
see to what extent it is possible to remove the cut-off. To do so, we will
renormalize the local coupling parameters so that the partition function as a
function of renormalized couplings remains finite as $\ep\to0$. In fact, the
quadratic nature of \Lb\ implies that the only renormalization needed is to
absorb into $a$ a divergent term, which is now a non-linear function of other
couplings, rather than proportional to $u^\mu$ as in the case of \Lp.

In Section 3, we proceed to analyse the string field action. The action is
determined from world-sheet two-point functions of boundary interactions and
their BRST transformation. Here we argue that to justify the formal
considerations of [\w], the BRST transformation of boundary operators should
not be modified by the presence of boundary interactions; otherwise the proof
that the antibracket is closed
does not go through.  This in fact is one reason that it is necessary
to begin the construction in a space of cut-off boundary interactions;
if the short distance behavior depends on the Lagrangian, the BRST
transformation
laws will also.

In trying to remove the cut-off, our main result is that the same
renormalization that makes the partition function finite also makes the string
field action finite. This is not obvious {\it a priori}.  However, after we
remove the cut-off, the fact that the short distance behavior of the local
theory \Lb\ depends on the boundary interaction comes back to haunt us in
the following way.  The string field action that we obtain is finite but
has an explicit dependence on $s$ in the following sense: the action
$S(t_0,\dots,t_s)$ constructed with one value of $s$ does not coincide,
if one sets $t_s=0$, with the action $S(t_0,\dots,t_{s-1})$ that one would
obtain starting from the outset with a smaller value of $s$.
This behavior is unpleasant, and we do not know the right interpretation.

\chapter{Partition Function}

The goal in this section is to solve the world-sheet matter theory defined by
$L_{\rm I}+L_{\rm B}$ on the disc by determining its matter partition
function. Since the action is quadratic, the partition function can simply
be expressed as a determinant of a corresponding Gaussian kernel, which in this
case is a differential operator on the circle, with a conventional (e.g.
$\zeta$-function) regularization. The approach here will be different, in that
the regulator will be introduced directly in the action by making $u(\th-\th')$
in \Lb\ a smooth function for a non-zero cut-off $\ep$, and the local boundary
interaction is recovered in the $\ep\to0$ limit. The partition function will be
determined by integrating a two-point Green's function, and will then be made
finite (as \ept) by a renormalization counter-term.

Let the unit disc be parametrized by polar coordinates $(r,\th)$ with the
boundary at $r=1$, as well as by complex coordinates $(z,\bar z)$ with
$z=r\e^{i\th}$. The variational principle applied to $\L+\Lb$ gives the
following boundary condition for $X_\mu$:
$$\pa_rX_\mu(\th) + \oint\d\th'u^\mn(\th-\th')X_\nu(\th') = 0. \eqn\bc$$
Here we have chosen $\eta^\mn=\de^\mn$ so that the spacetime indices may be
raised and lowered freely and later formulas may be simplified, but it is
obvious how to restore the Minkowski metric $\eta^\mn$ in what follows. The
exact Green's function $G_\mn(z,w)=\la X_\mu(z,\bar z)X_\nu(w,\bar w)\ra$
satisfying boundary condition \bc\ can be expressed as:
$$\eqalign{G_\mn(z,w)=&-\de_\mn\left(\ln|z-w|^2+\ln|1-z\bar w|^2\right)
+2(u_0^{-1})_\mn\cr
&-\sumkp \left({2u_k\over k(k+u_k)}(z\bar w)^k +
{2u_{-k}\over k(k+u_{-k})}(\bar zw)^k\right)_\mn ,\cr}\eqn\gf$$
where $u^\mn_k=u^\nm_{-k}=\oint\d\th\, u^\mn(\th)\,\emkt$ is the Fourier mode
of coupling function. When both positions are on the boundary, with
$z=\e^{i\th}$ and $w=\e^{i\th'}$, the Green's function becomes:
$$G_\mn(\th,\th')= \sumk\left({2\over |k|+u_k}\right)_\mn\,\,\e^{ik(\th-\th')}
\quad , \eqn\bgf$$
which determines the partition function $Z$ through the differential
equations:
$${\pa\over\pa\, u_k^\mn}\ln Z = -{1\over 16\pi^2}\oint\d\th\,\d\th'\,
\e^{ik(\th-\th')} \la X_\mu(\th)X_\nu(\th')\ra
= -\half \left({1\over |k|+u_{-k}}\right)_\mn\, .\eqn\pf$$

The partition function determined by \pf\ is valid for general (non-local)
boundary couplings. Consider now boundary interactions involving a finite
number of local couplings:
$$ L_{\rm B}= a + \factor\oint\d\th\sumn i^r\,t_r^\mn\,X_\mu\,\pa_\th^r
X_\nu(\th)\, .\eqn\Lbb$$
This corresponds to $u^\mn(\th)=\sumn\,(-i)^r\, t^\mn_r\,\de^{(r)}(\th)$, or
equivalently, $u^\mn_k=\sumn\,t^\mn_r k^r$. Now we introduce the short-distance
cut-off $\ep$ by taking:
$$u^\mn_k=\sumn\,t^\mn_r\,k^r\, \e^{-|k|\ep}\,.\eqn\uk$$
This is essentially a point-splitting regulator, and the particular form of the
cut-off dependence is chosen here for later convenience. Viewed as a function
of the coupling parameters $t_r^\mn=(-1)^r\,t_r^\nm$ and $a$, the partition
function now satisfies
$${\pa\over\pa\,t_r^\mn}\ln Z_s =-{\de_{r,0}\over 2}\left(u_0^{-1}\right)_\mn
-\half\sumkp \left( \left({k^r\,\ekep \over k+u_k}\right)_\nm
+(-1)^r \left({k^r\,\ekep \over k+u_k}\right)_\mn\right)\, ,\eqn\ppf$$
where the subscript $s$ indicates explicitly that the leading boundary coupling
is $t_s$. Using also ${\pa\over\pa\,a}\ln Z_s=-1$, we obtain
$$Z_s=(\det\,t_0)^{-1/2}\cdot\e^{-a}\cdot
\prod_{k=1}^\infty\det(1+k^{-1}\,u_k)^{-1/2}\, .\eqn\partition$$

The partition function \partition\ is divergent as \ept. The divergence comes
from the infinite sum in \ppf\ for $r=s$ and $r=s-1$. To make sense of the
partition function as \ept, one would like to view the ``bare'' couplings $t_r$
and $a$ as appropriate functions of ``renormalized'' couplings $t'_r$ and $a'$
and the cut-off $\ep$, such that the ``renormalized'' partition function,
$Z'_s(t',a';\ep)\equiv Z_s(t(t',a',\ep),a(t',a',\ep);\ep))$, now viewed as a
function of renormalized couplings, is finite in the limit $\ep\to0$. It is
easy to see that, for this quadratic theory, it is sufficient to take
$t_r=t'_r$ and $a=a(a',t'_r,\ep)=a'+\De a(t_r,\ep)$, and simply choose the
counter-term \da\ to
cancel the divergence in the partition function. To analyse this in detail, it
is convenient to consider the generating function:
$$\ln Z'_s(t_r,a';\ep)=-\half\tr\ln t_0-\tr\sumkp\ln(1+k^{-1}\,u_k)
-\De a-a'\,,\eqn\logz$$
where $u_k$ is given by \uk. With cut-off $\ep\not=0$, one may take derivatives
with respect to $t_r$ of the infinite sum in \logz, and show that the \ept\
divergent term is not a constant but depends on $t_s$ and $t_{s-1}$.
Furthermore, the dependence on $t_{s-1}$ is only linear. Thus in a ``minimal
subtraction'' scheme the counter-term $\De a$ may be chosen to depend only on
those two coupling parameters (for $s\geq 1$):
$$ \De a= -\tr\sumkp\left(\ln(1+t_s\,k^{s-1}\,\ekep) + {t_{s-1}\,k^{s-2}\,\ekep
\over 1+t_s\,k^{s-1}\,\ekep}\right) \,.\eqn\ct$$

Once the divergences are cancelled, one may take the $\ep\to0$ limit of \logz\
or \partition\ to obtain the ``renormalized'' generating function or partition
function. The generating function $\ln Z'_s$ so determined is exact and
generates arbitrary correlation functions of the (integrated) boundary
operators associated with the couplings $t_r, 0\leq r\leq s$. In particular all
these correlation functions are finite once $\ln Z'_s$ is made finite. The
correlation functions of other composite boundary operators are not generated
by $\ln Z'_s$ and those of operators with higher dimensions, such as
$X_\mu\,\pa_\th^r X_\nu(\th)$ for $r>s$, will still be divergent. (Correlation
functions of bulk operators can be computed using Wick's theorem from the exact
Green's function (2.2), and they have the usual short-distance behavior.)

It is worthwhile to consider special examples with $s=0,1$.
The $s=0$ case corresponds to quadratic tachyon considered in [\ww]. One finds
that the divergent part of the bare partition function is linear in $t_0$, and
the minimal counter-term \da\ can be chosen to be
$$ \De a =- {\rm Tr}(t_0)\sumkp{\ekep\over k}\, ,\quad s=0.\eqn\cttt$$
This is the same logarithmically divergent counter-term as used in the
normal-ordering prescription in [\ww]. With this counter-term, the renormalized
partition function $Z'_0$ can be seen to agree with that of [\ww].

For $s=1$, the second term in \ct\ is $\ln\ep$ divergent, while the first term
contains an $\epsilon^{-1}$ divergence and a finite term. Defining \da\ without
the finite term and using $t_r^\mn=-(-1)^r\,t_r^\nm$ to simplify further, one
finds
$$\De a=\ep^{-1}\sum_{m=0}^\infty {{\rm Tr}\,(t_1^{2m+2})\over (2m+2)^2}
+\ln\ep \sum_{m=0}^\infty {{\rm Tr}\,(t_0 t_1^{2m})\over 2m+1}\, ,\quad s=1.
\eqn\ctttt$$
This can also be compared with known results. Consider the case $t_0=0$, which
describes open string coupled to a background $U(1)$ gauge field with constant
field strength $t_1^\mn$. There are $X_\mu$ zero modes and the partition
function will be proportional to the spacetime volume, as indicated by the
first factor in \partition. The free energy (per unit volume) $\ln Z_1(t_1)$ is
defined by \logz\ without the first term. After substracting the counterm-term
\ctttt\ and setting $a'=0$ one finds
$$\ln Z'_1(t_1) = {1\over 4}{\rm Tr}\ln(1-t_1^2)\, ,\eqn\nouse$$
\REF\fradkin{E. Fradkin and A. A. Tseytlin, {\it Non-Linear Electrodynamics
from Quantized Strings},, Phys. Lett. {\bf B163} (1985) 123.}
\REF\nappi{A. Abouelsaood, C. G. Callan, C. R. Nappi, and S. A. Yost,
{\it Open Strings in Background Gauge Fields},
Nucl. Phys. {\bf B280} (1987) 599.}
in a complete agreement with earlier computations [\fradkin, \nappi]. It is
easy to see that the particular regulator and the minimal counter-term \ctttt\
used here is in fact equivalent to the $\zeta$-function regularization of
[\fradkin].

Although the minimal subtraction scheme is most natural and reproduces earlier
results for $s=0,1$ as remarked above, one may inquire whether other choices of
$\De a$ with different finite part might be more natural for general $s$.
Recall that in the theory of renormalization, ambiguities involving finite
counter-terms are removed by renormalization conditions and possibly some other
physical requirements such as symmetries. In the present problem, a natural
requirement is  the $s$ independence
of the partition function in the sense that
 $$Z'_s(a';t_0,...,t_{s-1},t_s)\vert_{t_s=0}=Z'_{s-1}(a';t_0,...,t_{s-1})\,,
\eqn\nat$$
as well as similar relations for the renormalized correlation functions in the
$Z_s$ and $Z_{s-1}$ theories. This requirement is satisfied at $s=1$ by the
minimal counter-terms \cttt\ and \ctttt, but is not satisfied by minimal
counter-term \ct\ for $s>1$. In fact $\ln Z'_s$ determined from \logz\ and \ct\
diverges as $t_s\to0$. Such singular behavior of the renormalized theory as
$t_s\to0$ is not unexpected in general, since $t_s$ (for $s>1$) is the leading
``unrenormalizable'' coupling which dominates the short-distance behavior.
Of course in a theory with a cut-off one may set $t_s=0$ without creating
divergences. Then the failure of the partition function to satisfy \nat\ is due
to the fact that the two limiting processes \ept\ and $t_s\to0$ do not commute.

The situation is simpler in the present case, as the bare partition function
\partition\ is naively $s$-independent and thus the $s$-dependence of the
renormalized partition function comes solely from the $s$-dependence of the
minimal counter-term \da. To satisfy \nat\ one simply needs to choose a
non-minimal but $s$-independent counter-term \da. There are many choices. The
simplest one is to take $\De a =-\tr\sumkp\ln(1+k^{-1}\,u_k)$, and the
renormalized partition function following from \logz\ is
$$Z_s=(\det\,t_0)^{-1/2}\cdot\e^{-a'}\, ,\eqn\simple$$
independent of all couplings except $t_0$, corresponding to normally ordering
$X\pa_\th^r X$ to have zero expectation value. But this does not give the $s=0$
and $s=1$ results expected from minimal counter-terms and from earlier
computations. There is a more complicated and but apparently natural choice
that does give the expected $s=0$ and $s=1$ results. It gives the following
$s$-independent renormalized generating function (with cut-off $\ep$ set to
zero):

$$\ln Z'_s=-{1\over 4}\tr\ln{t_0^2\over 1-t_1^2}
+\tr\sumkp\left(\ln\biggl(1-{t_0\over k+\sum_{r=0}^s t_r\,k^r}\biggr)
+{t_0\over k+\sum_{r=1}^s t_r\,k^r} \right) + a'\,.\eqn\part$$
Note that since all (integrated) correlation functions of the quadratic
boundary operators in the world-sheet action can be obtained by
differentiation of the generating function $\ln Z'_s$ with respect to the
coulplings, they will also be $s$-independent. More generally, arbitrary
correlation functions of interior or boundary operators may be obtained by
using Wick's theorem from the exact Green's function \gf, and they are
manifestly $s$-independent as well.

\chapter{String Field Action}
Let the boundary action be represented by $\oint\d\th\, b_{-1}O$, where $b$ is
the anti-ghost and $O$ has ghost number one. Introduce a basis $\{O_i\}$ of
ghost number one boundary operators, so that $O=\sum_i x^iO_i$. Here the
coupling constants $x_i$ may be viewed as parametrizing a point in the space of
world-sheet theories and thus also as parametrizing a string field
configuration. The string field action is defined in terms of world-sheet
correlation functions by [\w,\ww]:
$$\eqalign{ \d S &={1\over 2}\sum_{i,j} x^j\d x^i\oint {\d\th\,\d\th'}\,
\la O_i(\th) \,\,\{Q, O_j\}(\th')\ra \cr
&= {1\over 2}\oint {\d\th\,\d\th'}\,
\la \d O(\th) \,\,\{Q, O\}(\th')\ra\,,\cr}\eqn\sfa$$
where $\{Q,O_j\}$ is the BRST transformation of the boundary operator $O_j$.

Despite the boundary interaction, a conserved BRST current $(J_z, J_{\bar z})$
exists in the interior, given by that of standard closed string theory. The
action of BRST charge on a boundary operator $Q_j$ is
$$\{Q,O_j\} = \oint_C {\d z\over 2\pi i} J_z (z)\,O_j
- \oint_{\bar C} {\d\bar z\over 2\pi i} J_{\bar z}(\bar z)\,O_j\, ,\eqn\brst$$
where $C$ is a contour approaching the boundary of the disc and $\bar C$ is its
image under $z\to 1/z$.
Unrenormalizable boundary couplings ($t_r$ for $r>1$) in the absence of a
cut-off modify the short-distance behavior of the exact boundary Green's
function and thus the canonical structure between boundary operators, because
these boundary interactions dominate the (interior) kinetic term in the
world-sheet action. If the exact Green's function is used in extracting
short-distance behavior in \brst, the BRST transformation will cease to be
geometrical, e.g., $\{Q, X_\mu(\th)\}\not=c\,\pa_\th X_\mu(\th)$, where $c$ is
the $\th$ component of the ghost field on the boundary.

The BRST invariance of the antibracket defined on the space of world-sheet
theories in fact requires that the BRST transformation law should be not be
modified by the boundary interaction. This invariance is the statement that
$\d S$ defined in \sfa\ is indeed closed, and in proving
this crucial property[\w], one has made an important assumption that
$${\pa\over\pa x^j}\{Q, O_i\} =0\, , \eqn\cond$$
in notations described above.
\REF\se{S. Shatashvili and E. Verlinde, private communications}
(In a more direct argument that $\d\,S$ is indeed locally exact [\se], this
condition must also be used.)  Thus, to define the string field action, it
appears
that one must begin by working in a space of cut-off boundary interactions,
for which \cond\ holds, and then if one wishes one can try to remove the
cut-off.

For a general theory with quadratic interaction \Lb, we can take $O(\th)$
to be of the form
$$ O(\th)=c(\th)\left( {a\over 2\pi}  + {1\over 8\pi}
X_\mu(\th)\oint\d\th'\,u^\mn(\th-\th')
\,X_\nu(\th')\right)\equiv c(\th)V(\th)\, .\eqn\bo$$
The BRST transformation of $O(\th)$ can be computed from \brst\ with the
standard BRST current in closed string theory. In this computation, the matter
stress energy tensor will be contracted with the matter part of \bo~to extract
short-distance singularities. There will be no double $\la X X\ra$ contractions
here, since for a smooth (non-local) coupling function $u(\th-\th')$,
$X_\mu(\th)$ and $X_\nu(\th')$ in $O(\th)$ are located at separate points
on the boundary. For single contraction alone (and with the same short-distance
behavior of two-point Green's function as in the free theory), one has the
general formula
$\{Q, X_\mu(\th)\}=c\,\pa_\th X_\mu(\th)$. Applying this to \bo, one finds:
$$\{Q,O(\th)\}=cc'V(\th)-{1\over 8\pi} cX_\mu(\th)\oint\d\,
\th'\,u^\mn(\th-\th')c(\th')\pa_{\th'}X_\nu(\th')\,.\eqn\brs$$
Substituting \brs\ in \sfa\ and evaluating the ghost correlation function
according to:
$$\la c(\th'')c(\th)c(\th')\ra=2\left(\sin(\th-\th')+\sin(\th'-\th'')
+\sin(\th''-\th)\right)\,,\eqn\gc$$
one obtains:
$$\d S =\oint\d\th\,\d\th' \la\d V(\th)
\cdot \left(A(\th')+\cos(\th-\th')B(\th')+\sin(\th-\th')
C(\th')\right)\ra, \eqn\ds$$
where matter boundary operators $A$, $B$ and $C$ are given by
$$\eqalign{A(\th)&=-V(\th)+{1\over 8\pi}X_\mu(\th)\oint\d\,\th'\sin(\th-\th')
u^\mn(\th-\th')\pa X_\nu(\th')\,,\cr
B(\th)&={1\over 4\pi} X_\mu(\th)\oint\d\,\th'
u^\mn(\th-\th')\pa X_\nu(\th')\,,\cr
C(\th)&=0\,.\cr} \eqn\abc$$

Given the exact generating function $\ln Z$ for correlation functions of the
boundary operators, it is straightforward to evaluate $\d S$ from \ds.
To simplify the discussion here, we shall use a general result proven in [\ww]
to write down $S$ directly. This result states that if the boundary interaction
$V(\th)$ as introduced in \bo\ and the operator $A(\th)$ as in \ds\ have the
expansion in terms of a basis $\{V_i(\th)\}$ of matter operators:
$$V(\th)=\sum_i x^i V_i(\th)\,,\quad\quad
A(\th)=\sum_i \al^i V_i(\th)\,,\eqn\va$$
then the action is given by
$$S=\left(\sum_i\al^i{\pa\over\pa x^i} + g\right)Z(x^j)\,,\eqn\sz$$
where $Z$ is the world-sheet partition function and $g$ is a constant.
Here the set of couplings is $\{x^i\}=\{a,u^\mn_k\}$, with associated basis
$\{V_i(\th)\}=\{\pa_{x^i}V(\th)\}$. A simple calculation following these
definitions gives the corresponding functions
$\{\al^i\}\equiv\{\al,\al^\mn_k\}$:
$$\al=-a\,,\quad\quad\al^\mn_k=\half k(u^\mn_{k+1}-u^\mn_{k-1})
-u^\mn_k\,.\eqn\alpha$$
To determine the constant $g$, one may compute $S$ directly from \ds\ for
special couplings. Such a computation with $u_k^\mn=\de^\mn u_k$ yields $g=1$.
This, together with \sz\ and \alpha, gives the string field action for general
non-local quadratic boundary interaction:
$$S=\left(\sumk\left(\half k(u^\mn_{k+1}-u^\mn_{k-1})-u^\mn_k\right)
{\pa\over\pa u^\mn_k} + a + 1 \right) Z(u, a)\,, \eqn\action$$
where the partition function $Z$ is determined by \pf.

Note that since \action\ is derived for non-local boundary interactions, there
is no ultraviolet divergence. In particular, both formula \sz\ and the BRST
invariance $\d^2 S=0$ (which must be used to derive \sz [\ww]) can be
established rigorously. Now let the non-local coupling function $u(\th-\th')$
represent a set of local couplings $\{t_r, r=0,...,s\}$ with boundary cut-off
$\ep$ as introduced in Section 2. In Fourier modes, this is given by \uk,
$$u_k=\sumn\,t_r\,k^r\, \e^{-|k|\ep}\,.\eqn\ukk$$
Applying this change of variables to \action, one finds explicit
$\ep$-dependence, as well as implicit $\ep$-dependence through partition
function $Z$. To remove the cut-off, one must show that the action as a
function
of renormalized coupling has a finite limit as $\ep\to0$. Because of various
sources of $\ep\to0$ divergence in $S$, it appears at first almost impossible
that the counter-term in $a=a'+\De a$ used in Section 2 to make partition
function finite could also make $S$ finite. But this turns out to be the case,
as will be shown presently.

After the change of variables \ukk, the action \action\ becomes:
$$\eqalign{{S_s\over Z_s}
&=\half(\e^{-\ep}+\e^\ep)\,\sum_{r=1}^s\,\sum_{m=0}^{[{r-1\over 2}]}
c^r_{2m+1}\,t^\mn_r\,{\pa\,\ln Z_s\over\pa\,t^\mn_{r-2 m}}
-\sum_{r=0}^s t_r^\mn\,{\pa\,\ln Z_s\over\pa\,t^\mn_{r}}\,+a+1 \cr
& +(\e^{-\ep}-\e^\ep)\,\sum_{r=0}^s\,\sum_{m=0}^{[{r\over 2}]}
c^r_{2m}\,t^\mn_r \sumkp k^{r+1-2m}\,\ekep\,\,{\pa\,\ln Z_s\over\pa\,u^\mn_k}
\,, \cr} \eqn\soz$$
where $c^r_m={r!\over m!(r-m)!}$ and ${\pa\over\pa t_r}\ln Z$ is the
(integrated) bare Green's function given by \ppf. As $\ep\to0$, divergences
appear in $\sumkp$ in the second line. They also appear in ${\pa\over\pa
t_r}\ln Z$ for $r=s, s-1$, and in $a$, but those can all be attributed to the
counter-term \da\ when expressed in terms of the renormalized Green's function:
$${\pa\over\pa t_r}\ln Z'_s={\pa\over\pa t_r}\ln Z_s-{\pa\over\pa t_r}\De a
\,. \eqn\relation$$
For the purpose of demonstrating the cancellation of divergences, it is
sufficient to use the minimal form of the counter-term $\De a(t_{s-1},t_s)$
given by \ct.
To isolate various divergences in \soz, one uses following result:
$$\lim_{\ep\to0} \,\,\ep^p\cdot\sumkp{k^{s+p-3}\,\ekep\over
1+t_s k^{s-1}\,\ekep} =0\,, \quad\quad p=1,2,3,.... \eqn\math $$
Using \pf, \relation\ and \ct, and dropping terms that vanish as $\ep\to0$, one
finds:
$$\eqalign{{S_s\over Z_s}
&=\left(\sum_{r=1}^s \sum_{m=0}^{[{r-1\over 2}]}
c^r_{2m+1}\,t^\mn_r {\pa\,\ln Z'_s\over\pa\,t^\mn_{r-2 m}}
-\sum_{r=0}^s t_r^\mn {\pa\,\ln Z'_s\over\pa\,t^\mn_{r}} +a'+1\right) \cr
&+ \ep\cdot{\rm Tr}\sumkp{t_s\,k^{s}\,\ekep\over 1+ t_s\, k^{s-1}\,\ekep}\cr
&+\left((s-1)\,t_s^\mn{\pa\De a\over\pa\,t^\mn_s}+(s-2)\,t_{s-1}^\mn
{\pa\De a\over\pa\,t^\mn_{s-1}} +\De a \right)\,. \cr}\eqn\sozz$$

The first term in \sozz\ is finite, since it involves renormalized quantities.
The second term in \sozz, coming from the second line in \soz, would not be
present if the action $S$ was constructed directly in terms of local couplings
without the cut-off. This term has a non-trivial contribution even as \ept\
because the infinite sum $\sumkp$ diverges as $\epsilon^{-2}$. The third term
is also divergent, and is a function of $t_{s-1}$ and $t_s$ only for minimal
counter-term \da\ given by \ct. Using \ct\ one finds that the
$t_{s-1}$-dependent term is in fact finite (for $s>1$) in the limit $\ep\to0$,
and is given by:
$$f_1(t_{s-1},t_s)=(s-1)\,{\rm Tr}
\sumkp{t_{s-1}\,k^{s-2}\over(1+t_s\,k^{s-1})^2} \,.\eqn\f$$
The rest of the third term in \sozz\ is a function of $t_s$ and is divergent as
$\ep\to0$. But its divergent part cancels precisely that of the second term in
\sozz. This remarkable cancellation can be summarized by the following
equation:
$$\eqalign{
&\ep\cdot{\rm Tr}\sumkp{t_s\,k^{s}\,\ekep\over 1+ t_{s}\, k^{s-1}\,\ekep}
-{\rm Tr}\sumkp\ln\left( 1+ t_s\, k^{s-1}\,\ekep\right) \cr
&\quad\quad\quad-(s-1)\,{\rm Tr}\sumkp{t_s\,k^{s-1}\,\ekep\over 1+ t_s\,
k^{s-1}\,\ekep} \cr
=& {\rm Tr}\sum_{m=1}^\infty {1 \over m}\, B_{m(s-1)}\,(t_s)^{m} +O(\ep)
\equiv f_2(t_s) + O(\ep) \, ,\cr}\eqn\finite$$
where $B_n$ are the Bernoulli numbers. The finiteness of \finite\ as \ept\ can
be seen easily by replacing $\sumkp$ with $\int_1^\infty \d k$ and integrating
by parts. To determine the finite part $f_2$, we expand each term in \finite\
in power series of $t_s$ (which is absolutely convergent for
$t_s<(\e\cdot\ep)^{s-1}/(s-1)^{s-1}$), and perform the sum $\sumkp$. The result
is a polynomial in $\ep$ with $f_2$ as the $\ep^0$ term (the negative powers of
$\ep$ cancel among the three terms in \finite). The cancellation of divergences
for $s>1$ means that the string field
action $S$ has a well-defined $\ep\to0$ limit given by:
$$S_s=\left(\sum_{r=1}^s \sum_{m=0}^{[{r-1\over 2}]}c^r_{2m+1}\,t^\mn_r
{\pa \over\pa\,t^\mn_{r-2 m}} -\sum_{r=0}^s t_r^\mn {\pa \over\pa\,t^\mn_{r}}
+f_1+f_2 +a'+1\right)Z'_s(t_r,a')\,,\quad s>1\,,\eqn\final$$
where $Z'_s(t_r,a')\equiv Z_s(t_r,a'+\De a)$ is the renormalized partition
function \logz.

The analysis of \sozz\ for $s=0,1$ is slightly different. For $s=0$, the third
term in \sozz\ is zero, since $\De a$ given in \cttt\ is linear in $t_0$. The
second term in \sozz\ is in fact finite as \ept\ and gives ${\rm Tr}(t_0)$. One
obtains the action for the quadratic tachyon:
$$S_0=\left({\rm Tr}(t_0) - t_0^\mn {\pa \over\pa\,t^\mn_0}
+a'+1 \right)\,Z'_0\,,\quad\quad s=0.\eqn\actionnn$$
This agrees completely with the result of [\ww]. In [\ww] there is no boundary
cut-off and the first term in \actionnn\ can be traced back to the double
contraction in computing the BRST transformation $\{Q,O\}$ of the boundary
operator. In the present formulation, there is no double contraction in
$\{Q,O\}$ but the first term in \actionnn\ does appear correctly after the
cut-off is removed. Therefore the agreement of \actionnn\ with [\ww] is highly
non-trivial and provides an important consistency check.
For $s=1$ one finds, using \ctttt, that divergences cancel between the second
and third terms in \sozz\ also. But there are now finite contributions from
both terms as $\ep\to0$. Together they give
${\rm Tr}\,t_0\,(1-t_1^2)^{-1}$, and the action is,
$$S_1=\left({\rm Tr}{t_0\over 1-t_1^2} - t_0^\mn {\pa \over\pa\,t^\mn_0}
+a'+1 \right)\,Z'_1\,,\quad\quad s=1.\eqn\actionn$$

{}From the computations described above, it appears miraculous that the
counter-term $\De a$ that makes the matter partition function finite in Section
2 cancels all divergences in the action \soz. To see that this cancellation is
non-trivial, we first note that the action is determined from the {\it bare}
two-point correlation functions of the world-sheet
theory; there is no counter-term or renormalization other than those present
in the world-sheet Lagrangian.
When expressed in terms of renormalized couplings, the
action \soz\ contains divergences (the third term in \sozz) coming from the
world-sheet counter-term \da\ . The additional divergence (the second term in
\soz\ or \sozz) can be traced back to the BRST transformation of the leading
unrenormalizable boundary operator $X\pa^s_\th X$. Because
$\{Q, X_\mu(\th)\}=c\,\pa_\th X_\mu(\th)$, the operator $X\pa^{s+1}_\th X$ with
one higher dimension may be produced by BRST transformation, and its two-point
function will contribute to the string field action. More precisely, as can
be seen from the second term in $A(\th)$, this contribution is multiplied by
the short-distance cut-off ($\sin(\th-\th')\sim\ep$) and it corresponds to
precisely the divergent second term in \soz\ and \sozz. From the world-sheet
point of view,
correlation functions of boundary operators $X\pa^{s+1}_\th X$ can be made
finite only if the correponding coupling and its renormalization is introduced.
Then by induction one would seem to require an infinite number of couplings and
their renormalization to obtain a finite string field action as \ept.
What actually happens is a complete concellation of divergence with only
renormalization of a finite number of local couplings.

Other choices of counter-term \da\ with different finite part will also give a
finite and cut-off independent string field action \final, but with possibly
different finite term $f_1+f_2$. For the minimal counter-term, the action
\final\ is not $s$-independent, as the $t_s\to0$ limit is singular. As shown in
Section two, a non-minimal but $s$-independent counter-term can been
constructed
to give a $s$-independent renormalized partition function. Unfortunately these
$s$-independent counter-terms still do not give an $s$-independent action. This
is because of the explicit $s$-dependence of the second term in \sozz; the two
limits \ept\ and $t_s\to0$ do not commute for this term. The failure of the
action to be $s$-independent may present difficulties to the notion that the
space of open string theories should be represented by local world-sheet
boundary interactions. It could be that, in fact (as has been suggested
to us by D. Gross), the cut-off should only be
removed if one is sufficiently close to classical solutions.  (For the specific
boundary interactions considered in this paper, the massive space-time fields
are taken at zero momentum and so are far off-shell.)  However, we do not know
any systematic way to proceed -- or
to achieve background independence -- along those lines.  We do believe
that the problem of interpreting or dealing with the unrenormalizable
world-sheet interactions is one of the main obstructions to progress
in string theory.

\ack
We are  grateful to Michael Douglas, David Gross, Samson Shatashvili and Erik
Verlinde for helpful discussions.
\refout

\vfill\eject\end